\documentclass{PoS}

\usepackage{amsmath}
\usepackage{amssymb}
\usepackage{epsfig}

\newcommand{\ep}{\varepsilon}

\newcommand{\NN}{\mathbb{N}}
\newcommand{\ZZ}{\mathbb{Z}}

\newcommand{\RR}{\mathbb{R}}
\newcommand{\sign}{\text{sign}}

\title{{\footnotesize DESY 14--122, DO-TH 14/13, SFB/CPP-14-039, LPN14-086}\\
Recent Symbolic Summation Methods to Solve Coupled Systems of
Differential and Difference Equations}

\ShortTitle{Solving Coupled Systems of
Differential and Difference Equations}

\author{\speaker{Carsten Schneider}\thanks{This work was supported in part by
DFG Sonderforschungsbereich 
Transregio 9, Computergest\"utzte Theoretische Teilchenphysik, 
the Austrian Science Fund (FWF) grants P20347-N18 and SFB F50 (F5009-N15), and
the European Commission through 
contract PITN-GA-2010-264564 ({LHCPhenoNet}) and PITN-GA-2012-316704
({HIGGSTOOLS}).}\\
        Research Institute for Symbolic Computation (RISC)\\
        Johannes Kepler University, Altenbergerstra\ss{}e 69, A-4040 Linz,
Austria\\
        E-mail: \email{Carsten.Schneider@risc.jku.at}}

\author{Johannes Bl\"umlein and Abilio de Freitas\\
        Deutsches Elektronen--Synchrotron, DESY,\\
Platanenallee 6, D--15738 Zeuthen, Germany\\
        E-mail: \email{johannes.bluemlein@desy.de,abilio.de.freitas@desy.de}}

\abstract{We outline a new algorithm to solve coupled systems of differential
equations in one continuous variable $x$ (resp.\
coupled difference equations in one discrete variable $N$) depending on a small
parameter $\ep$: given such a system and given sufficiently many initial values,
we can determine the first coefficients of the Laurent-series solutions in $\ep$
if they are expressible in terms of indefinite nested sums and
products. This systematic approach is based on symbolic summation algorithms in
the context of difference rings/fields and uncoupling algorithms.  The proposed
method gives rise to new interesting applications in connection with integration
by parts (IBP) methods. As an illustrative example, we will demonstrate how one
can
calculate the $\ep$-expansion of a ladder graph with 6 massive fermion lines.}

\FullConference{ Loops and Legs in Quantum Field Theory - LL 2014,\\
		 27 April - 2 May 2014 \\
		 Weimar, Germany }

\begin{document}

\section{Introduction}

Symbolic summation in the setting of difference fields and
rings~\cite{Karr:81,Schneider:01,Bron:00,Schneider:05a,Schneider:08c,
Schneider:10a, Schneider:10b,
Schneider:10c,Schneider:13b} provides a general toolbox in form of the
Mathematica package \texttt{Sigma}~\cite{SIG1,SIG2} to
simplify definite multi-sums to expressions in terms of indefinite nested sums
and products. This function domain covers as special cases harmonic
sums~\cite{Blumlein:1998if,Vermaseren:1998uu}
\begin{equation}\label{Equ:HarmonicSums}
S_{a_1,\dots,a_k}(N)= \sum_{i_1=1}^N\frac{\sign(a_1)^{i_1}}{i_1^{|a_1|}}
\sum_{i_2=1}^{i_1}\frac{\sign(a_2)^{i_2}} {i_2^{|a_2|}}\dots
\sum_{i_k=1}^{i_{k-1}}\frac{\sign(a_k)^{i_k}}{i_k^{|a_k|}},\quad
a_i\in\ZZ\setminus\{0\},
\end{equation}
generalized harmonic sums~\cite{Moch:2001zr,Ablinger:2013cf}, 
cyclotomic sums~\cite{Ablinger:2011te} or
nested binomial
sums~\cite{Fleischer:1998nb,Davydychev:2003mv,Weinzierl:2004bn,Ablinger:2014bra} .

In the last years this technology proved to be useful to evaluate non-trivial
Feynman integrals in the context of QCD.
Namely, as worked out in~\cite{Blumlein:2010zv,Weinzierl:13} a big
class of integrals in terms of the dimensional parameter $\varepsilon=D-4$
and a discrete Mellin parameter $N$ can be written in the following form:
\begin{equation}\label{Equ:MultiSum}
F(N)=\sum_{k_1=l_1}^{L_1(N)} ... \sum_{k_v=l_v}^{L_v(N,k_1, ..., k_{v-1})}
f(\varepsilon,N,k_1,\dots,k_v)
\end{equation}
\noindent where $L_i(N,k_1,\dots,k_{v-1})$ stands for an integer linear relation
in the variables $N,k_1,\dots,k_{v-1}$ or is $\infty$ and
$f(\varepsilon,N,k_1,\dots,k_v)$ is a linear combination of proper
hypergeometric
sequences given in terms of $\Gamma$-functions with arguments in terms of
integer linear relation in the integer parameters $N,k_1,\dots,k_{v-1}$ and
$\varepsilon$ might occur linearly in the form $r\,\varepsilon$ with $r$ being a
rational number.

Given such a multi-sum $F(N)$, the main task is to compute the
first coefficients $F_i(N)$ of the Laurent-series expansion
\begin{equation}\label{Equ:FExpansion}
F(N)\stackrel{?}{=}F_{\lambda}(N)\ep^{\lambda}+F_{\lambda+1}(N)\ep^{\lambda+1}
+\dots, \quad\lambda\in\ZZ
\end{equation}
in terms of special functions such as (generalized) (cyclotomic) harmonic sums
and nested binomial sums mentioned above. To get this representation,
the following two tactics are of interest.

\smallskip

\noindent\textbf{Tactic 1:} Compute the
coefficients of the $\ep$-expansion of the summand 
\begin{equation}\label{Equ:FSummandExpansion}
f(\varepsilon,N,k_1,\dots,k_v)=f_{\lambda}(N,k_1,\dots,k_v)\ep^{\lambda}+f_{
\lambda+1 }
(N,k_1,\dots,k_v)\ep^{\lambda+1}\dots
\end{equation}
by formulas such as Eq.~(1.4) in~\cite{Ablinger:2012LL} and arrive at
a linear combination of hypergeometric terms (free of $\ep$) multiplied with
(cyclotomic) harmonic sums. Finally, if the interchange of
the summation signs and differentiation w.r.t.\ $\ep$ (which we applied
to get the summand expansion) is valid, we end up at multi-sum
representations for the coefficients of the expansion~\eqref{Equ:FExpansion}:
$$F_i(N)=\sum_{k_1=l_1}^{L_1(N)} ... \sum_{k_v=l_v}^{L_v(N,k_1,
..., k_{v-1})}
f_i(N,k_1,\dots,k_v).$$
Note that exactly at this point
our symbolic summation toolbox in form of the Mathematica package
\texttt{EvaluateMultiSums}~\cite{Ablinger:2010pb,SIG2} (based on the
difference field/ring algorithms in \texttt{Sigma}) can be activated: it tries
to
transform the found multi-sums completely automatically to the desired form
in terms of indefinite nested sums and products. If infinite summation bounds
occur one needs in addition the Mathematica package
\texttt{HarmonicSums}~\cite{Ablinger:2011te,Ablinger:2013hcp,Ablinger:2013cf}
which provides the necessary asymptotic expansions to treat limit computations. 

\smallskip

\noindent\textbf{Tactic 2:} A different approach is as follows. Hunt for a
recurrence
\begin{multline}\label{Equ:epRec}
a_0(\ep,N)F(N)+a_1(\ep,N)F(N+1)+\dots+a_d(\ep,
N)F(N+d)\\
=h_{\lambda}(N)\ep^{\lambda}+h_{\lambda+1}(N)\ep^{\lambda+1}+h_{\lambda+2}
(N)\ep^{\lambda+2}+\dots
\end{multline}
\normalsize
of order $d\in\NN$ with polynomials $a_i(\ep,N)$ in the variables
$\ep$ and $N$. In particular, we require that the inhomogeneous part is given in
expanded form where the coefficients $h_i(N)$ are expressions in terms
of indefinite nested sums and
products. For the different methods and algorithms to compute such recurrences
we refer
to~\cite{Blumlein:2010zv,Ablinger:2012LL} and references therein. Here we
emphasize the
following~\cite{Blumlein:2010zv}: Given such a recurrence and suppose that we
are given the initial values $F(n)$
for $n=1,\dots,d$ expanded high enough, then we are in business to obtain the
all-N solution using \texttt{Sigma}: we can determine the first coefficients of
the
expansion~\eqref{Equ:FExpansion} whenever they are expressible in terms of
indefinite nested sums and products.

\smallskip

These tactics (in particular the first variant with \texttt{EvaluateMultiSums})
turned out to be instrumental to
evaluate two and three loop massive
integrals in~\cite{Bierenbaum:2008yu,Ablinger:2010ty,Ablinger:2012qm}.
Another interesting feature is to crunch the occurring sums with the Mathematica
package
\texttt{SumProduction}~\cite{Blumlein:2012hg} to basis sums (master sums) such
that no relations (in particular, no contiguous relations)
occur among them. Then our summation tools are only applied to a few
remaining sums. This enabled us to handle many additional problems such
as outlined in Refs.~\cite{Blumlein:2012vq,Ablinger:2014uka} and is currently used for ongoing
calculations. For the interplay of all these packages and their 
features we refer to~\cite{Schneider:2013zna}.

\smallskip

For our current calculations, we continue this strategy of
compactification by another component.  Namely, in order to
calculate massive 3-loop operator matrix elements~\cite{Ablinger:2014lka,Ablinger:2014vwa,Ablinger:14PS} we 
used integration by parts
(IBP) technology~\cite{Tkachov:1981wb,Chetyrkin:1981qh,Laporta:1996mq}, more
precisely, the powerful \texttt{C++}--code
\texttt{Reduze\!~\!2}~\cite{Studerus:2009ye,vonManteuffel:2012np} was used, to
reduce the input expression to a
reasonable number of master integrals. Then using our symbolic summation tools
(Tactic 1), we could expand the given master integrals in terms of harmonic
sums. Given these building blocks we could derive the expansion of the
complete input expression. 

However, in the
most recent calculations integrals occur that seem too hard for direct
calculations. This
pushed us forward to another aspect. We use the well known
fact that the master integrals are related
to each other~\cite{Laporta:2001dd}: together they form a hierarchically ordered
coupled system of
differential (resp.\ difference) equations. 

This article provides a new component to utilize this property by means
of symbolic summation (Section~\ref{Sec:SymbolicSummation}) and uncoupling
algorithms~\cite{Gerhold:02}. In Section~\ref{Sec:SolveCoupledSys} we
will present an algorithm that solves such coupled differential equations
in one
continuous variable $x$ (resp.\ coupled difference equations in a discrete
variable $N$). More precisely, we can derive the first coefficients of the
$\ep$-expansion of the master integrals using as input this coupled system and a
certain amount of initial values. A summary will be given in
Section~\ref{Sec:Conclusion}.

\newpage

\section{The backbone of our solver:
difference field/ring algorithms}\label{Sec:SymbolicSummation}

Subsequently, we work out the essential summation paradigms to treat the
two tactics presented above. The underlying ideas will
be demonstrated by tackling the following sum
\begin{equation}\label{Equ:SingleSum}
F(N)=\sum_{k=1}^{N}\underbrace{(-1)^k
e^{-\frac{3 \ep \gamma }{2}} \Gamma(-1-\frac{3
\ep}{2}\big)!B\big(2+k,\frac{\ep}{2}\big) B(-\ep+k,-\ep)
B\big(1-\frac{\ep}{2}+k,1+\frac{\ep}{2}\big)
\binom{N}{k}}_{f(N,k)}
\end{equation}
with Euler's $\gamma$ constant and where
$B(a,b)=\frac{\Gamma(a)\Gamma(b)}{\Gamma(a+b)}$ denotes the Beta-function.

\subsection{Tactic 1: Expand under the summation sign and simplify the
coefficients}

As worked out in the introduction, we first compute the first coefficients
$f_i(N,k)$ of the $\ep$-expansion~\eqref{Equ:FSummandExpansion} of the summand
$f(N,k)$; here we have $\lambda=-3$. Then we get the
coefficients
$F_i(N)=\sum_{k=1}^N f_i(N,k)$ of the
$\ep$-expansion~\eqref{Equ:FSummandExpansion}. E.g., for $i=-1$ 
we get the single pole term
$$F_{-1}(N)=\sum_{k=1}^N (-1)^{k+1}\binom{N}{k}
\Big(\frac{(2+3 k) \big(-2+3
k+7 k^2+3 k^3\big)}{3 k^2 (1+k)^3}+\frac{2 S_2(k)}{1+k}+\frac{\zeta_2}{2
(1+k)}\Big)$$
with $\zeta_a=\sum_{i=1}^{\infty}\frac1{i^a}$. In order to simplify this sum
we
compute a recurrence relation using the summation package \texttt{Sigma}:
\footnotesize
\begin{multline}
\big(16 N^3+144 N^2+413 N+384\big) (N+1)^2 F_{-1}(N)
-(N+2) (2 N+5) \big(16 N^3+112 N^2+221 N+113\big) F_{-1}(N+1)\\
+(N+3)^2 \big(16 N^3+96 N^2+173 N+99\big) F_{-1}(N+2)
=\tfrac{\zeta_2\big(4 N^2+21 N+29\big)}{2}+\tfrac{-64 N^5-500 N^4-1133 N^3+203
N^2+3516 N+3090}{3 (N+2) (N+3)}.
\end{multline}
\normalsize

\noindent\textit{Remark.} The underlying difference field
algorithms~\cite{Schneider:01,Schneider:08c,Schneider:10a,
Schneider:10b,Schneider:13b} are based on
Zeilberger's creative
telescoping paradigm~\cite{Zeilberger:91}. In general, the input is a definite
sum $\sum_k f(N,k)$ where the summand $f$
may consist of indefinite nested sums and products w.r.t.\ the summation
variable $k$ and where the occurring objects in $f$ might depend on the
parameter $N$ (or even further parameters). Note that the algorithms provide
also a proof
certificate that shows the correctness of the
recurrence.

\medskip

\noindent Now we activate \texttt{Sigma}'s recurrence solver, which can
handle
the following problem~\cite{Petkov:92,Abramov:94,Schneider:01}.

\medskip

\noindent\textbf{Problem REC:}\label{Problem:REC} GIVEN polynomials
$a_0(N),\dots,a_d(N)$ in $N$
and an expression $h(N)$ in terms of indefinite nested sums and products (such
as harmonic sums, binomial nested sums, etc.).\\ 
FIND all solutions of the
linear recurrence
$$a_0(N)F(N)+\dots+a_d(N)F(N+d)=h(N)$$
that are expressible
in terms of indefinite nested
sums and products. 
\medskip

\noindent In our particular instance, we find the solutions
\begin{align*}
L=\big\{&c_1\frac{1-4 N}{N+1}+
c_2\,\Big(\frac{-14 N-13}{(N+1)^2}
+\frac{(4 N-1) S_1(N)}{N+1}\Big)
+ \frac{(1-4 N) S_1(N)^2}{6 (N+1)}\\
&+\frac{(14 N+13) S_1(N)}{3 (N+1)^2}+\frac{175
N^2+334 N+155}{12 (N+1)^3}
+\frac{(1-4 N) S_2(N)}{6 (N+1)}+\frac{\zeta_2}{8 (N+1)}
|c_1,c_2\in\RR\}.
\end{align*}
\noindent Since the solution set is completely determined (note that
we found two
linearly independent solutions of the homogeneous version and one particular
solution of the recurrence itself), it follows that 
$F_{-1}(N)\in L$. The first two initial values $N=1,2$ determine uniquely 
$c_1=\tfrac1{12}-\tfrac{\zeta_2}{8}$ and $c_2=1$. Summarizing,
we discovered
(together with a rigorous proof) that
\begin{align*}
F_{-1}(N)=&\big(\frac{1}{12}
-\frac{1}{8}\zeta_2\big)\,\frac{1-4 N}{N+1}+
\frac{-14 N-13}{(N+1)^2}+\frac{(4 N-1) S_1(N)}{N+1}
+ \frac{(1-4 N) S_1(N)^2}{6 (N+1)}\\
&+\frac{(14 N+13) S_1(N)}{3 (N+1)^2}+\frac{175 N^2+334 N+155}{12 (N+1)^3}
+\frac{(1-4 N) S_2(N)}{6 (N+1)}+\frac{\zeta_2}{8 (N+1)}.
\end{align*}
We remark that the package \texttt{EvaluateMultiSums} combines all the
available features of \texttt{Sigma} yielding a powerful function to obtain such
simplifications in terms of indefinite nested sums and products completely
automatically.

\subsection{Tactic 2: Extract the expansion from a recurrence}

For the second tactic we need a recurrence~\eqref{Equ:epRec} for our
sum~\eqref{Equ:SingleSum}. Using \texttt{Sigma} we get
\begin{multline*}
2 (N+1)^2 F(N)+\big(3 \ep^2+3 \ep N+9 \ep-4 N^2-12 N-8\big) F(N+1)\\
-(2 \ep-N-1) (\ep+2 N+6) F(N+2)
=0\ep^{-3}-\tfrac{16}{3}\ep^{-2}+\tfrac{40}{3}\ep^{-1}-\big(2
\zeta_2-\tfrac{68}{3})\ep^{0}+\dots.
\end{multline*}
Together with initial values for $N=1$ and $N=2$  
\begin{align*}
F(1)&=\tfrac{2}{3}\ep^{-3}-\tfrac{11}{6}\ep^{-2}+\big(\tfrac{\zeta_2}{4}+\tfrac{
79 } {
24}\big)\ep^{-1}+\dots,&
F(2)&=\tfrac{8}{9}\ep^{-3}-\tfrac{73}{27}\ep^{-2}+\big(\tfrac{\zeta_2}{3}+\tfrac
{
1415}{324}\big)\ep^{-1}+\dots
\end{align*}
we are now
in the position to calculate the first coefficients of the $\ep$-expansion with
\texttt{Sigma}:
\begin{align*}
F&(N)=\tfrac{4 N}{3 (N+1)}\ep^{-3} -\Big(\tfrac{2 (2 N+1)}{3
(N+1)}S_1(N)+\tfrac{2 N (2 N+3)}{3 (N+1)^2}\Big)\ep^{-2}\\
&\Big(\tfrac{(1-4 N) }{6 (N+1)}S_1(N)^2-\tfrac{N \big(N^2-2\big)}{3
(N+1)^3}+\tfrac{(3 N+2) (4 N+5)}{3 (N+1)^2}S_1(N)+\tfrac{(1-4 N)}{6
(N+1)}S_2(N)+\tfrac{N \zeta_2}{2 (N+1)}\Big)\ep^{-1}+\dots
\end{align*}
In general, suppose we are given a recurrence~\eqref{Equ:epRec} with
polynomial coefficients
$a_i(N)$ (not all $a_i$ being the zero-polynomial) and expressions $h_i(N)$ in
terms of indefinite nested sums and products; furthermore assume we are given
the expansion of $F(i)$
for $i=1,\dots,d$ up the the order $\ep^r$. Then we can decide if the
coefficients of
the expansion~\eqref{Equ:FExpansion} up to order $r$ can be expressed in
terms of indefinite nested sums and products. 

Here the general idea is as follows. Plug in the generic
solution~\eqref{Equ:FExpansion} into~\eqref{Equ:epRec}:
\begin{equation}\label{Equ:AnsatzSolver}
\begin{split}
&a_0(\ep,N)\Big[F_{\lambda}(N)\ep^{\lambda}+F_{\lambda+1}(N)\ep^{\lambda+1}
+F_{\lambda+2}(N)\ep^{\lambda+2}\dots\Big]\\
+&a_1(\ep,N)\Big[F_{\lambda}(N+1)\ep^{\lambda}+F_{\lambda+1}(N+1)\ep^{\lambda+1}
+F_{\lambda+2}(N+1)\ep^{\lambda+2}\big]+\dots+\\
+&a_d(\ep,N)\Big[F_{\lambda}(N+d)\ep^{\lambda}+F_{\lambda+1}(N+d)\ep^{\lambda+1}
+F_{\lambda+2}(N+d)\ep^{\lambda+2}+\dots\Big]\\
&\hspace*{6cm}
=h_{\lambda}(N)\ep^{\lambda}+h_{\lambda+1}(N)\ep^{\lambda+1}+h_{\lambda+2}
(N)\ep^{\lambda+2}+\dots
\end{split}
\end{equation}
Two Laurent series agree if they agree term-wise, in particular the term
with lowest order must agree. This gives the constraint\footnote{
We auppose
that $a_i(0,N)\neq0$ for all $i$; otherwise divide through
$\ep$ several times which amounts to decrease $\lambda$.}
$$a_0(0,N)F_{\lambda}(N)+a_1(0,N)F_{\lambda}(N+1)+\dots+a_{d}(0,N)F_{
\lambda}(N+d)=h_{\lambda}(N).$$
Now we activate the recurrence solver (see Problem REC on
page~\pageref{Problem:REC}) and calculate with \texttt{Sigma} all solutions
that are expressible in terms of indefinite nested sums and products.
Thus together with the initial values for $F_{\lambda}(1),\dots,F_{\lambda}(d)$
we can decide if $F_{\lambda}(N)$ can be expressed in terms of indefinite nested
sums and products. If this fails, our algorithm stops. Otherwise, we take
the found representation of $F_{\lambda}(N)$ in terms of indefinite nested
sums and products and plug it into~\eqref{Equ:AnsatzSolver}.
Shuffling the inserted expressions to the right hand side gives
\begin{equation*}
\begin{split}
&a_0(\ep,N)\Big[F_{\lambda+1}(N)\ep^{\lambda+1}
+F_{\lambda+2}(N)\ep^{\lambda+2}\dots\Big]\\
+&a_1(\ep,N)\Big[F_{\lambda+1}(N+1)\ep^{\lambda+1}
+F_{\lambda+2}(N+1)\ep^{\lambda+2}\big]+\dots+\\
+&a_d(\ep,N)\Big[F_{\lambda+1}(N+d)\ep^{\lambda+1}
+F_{\lambda+2}(N+d)\ep^{\lambda+2}+\dots\Big]
=h'_{\lambda+1}(N)\ep^{\lambda+1}+h'_{\lambda+2}
(N)\ep^{\lambda+2}+\dots
\end{split}
\end{equation*}
where the $h'_i(N)$ are updated expressions in terms of indefinite nested sums
and products. By construction the $\ep^{\lambda}$ contribution is removed and we
can divide the whole equation by $\ep$. Thus we can repeat the procedure where
$F_{\lambda+1}$ (instead of $F_{\lambda}$) plays the role of the lowest term.

\section{A challenging diagram and a new algorithm to solve
coupled systems}\label{Sec:SolveCoupledSys}

We want to calculate the $\ep$-expansion of the ladder
graph with 6 massive 
fermion lines\footnote{The following graph has been drawn using {\tt Axodraw} \cite{Vermaseren:1994je}.}:
\begin{equation}\label{Equ:Diagram}
D_4(N)=\begin{minipage}{3cm}
\epsfig{figure=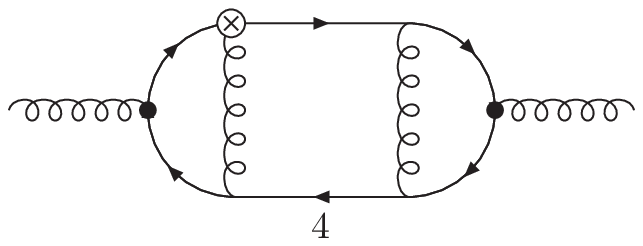,width=1\linewidth}
\end{minipage}\stackrel{?}{=}F_{-3}(N)\ep^{-3}+F_{-2}(N)\ep^{-2}+F_{-1}(N)\ep^
{-1}+F_{0}
(N)\ep^{0}+\dots
\end{equation}
For scalar diagrams of the same class~\cite{Ablinger:2012qm} we succeeded in
calculating the
$\ep$-expansion following Tactic~1 of Section~\ref{Sec:SymbolicSummation}.
But for the diagram in question we failed with this tactic so far.
As it turned out, a clever extension~\cite{Ablinger:2014yaa} of
Brown's hyperlogarithm algorithm~\cite{Brown:2008um} was successful to obtain
in~\cite{Ablinger:2012qm} the
scalar version of~\eqref{Equ:Diagram}, i.e. for diagrams with a numerator function equal to one, with
$\lim_{\ep\to0} D_4(N)=F_0(N)$.
However, if one wants to consider the complete physical diagram, this
method does not apply since poles in $\varepsilon$ occur in almost all
integrals. Subsequently, 
we present a new strategy
that can tackle such diagrams in a rather natural way.

Let us consider the generating function (formal power series) of
$D_4(N)$, i.e.,
$$\hat{D}_4(x)=\sum_{N=0}^{\infty}D_4(N)x^N.$$
Then by using refined IBP methods, i.e., by
\texttt{Reduze\!~\!2}~\cite{Studerus:2009ye,vonManteuffel:2012np,
Ablinger:2014vwa} we obtain
the expression 
\begin{align}\label{Equ:D4XSpace}
\sum_{N=0}^{\infty}D_4(N)x^N
=&\sum_{i=1}^{52}\square\hat{B}_i(x)+\sum_{i=1}^{15}\square
\hat{I}_1(x)
\end{align}
with the master integrals $\hat{I}_i$ and $\hat{B}_i$ and with large
coefficients
$\square$ in terms of rational functions in $\ep$ and $x$ that are not printed
out explicitly. As it turns out, $\hat{B}_1(x),\dots,\hat{B}_{52}(x)$ can be
determined directly with sophisticated Mellin-Barnes techniques and our
summation tools from above. E.g., for
$\hat{B}_1(x)=\sum_{N=0}^{\infty} B_1(N)x^N$
the integral $B_1(N)$ can be written in the compact sum
representation~\eqref{Equ:SingleSum} for which we carried out the
$\ep$-expansion as a concrete example.

Since $\hat{I}_1(x),\dots,\hat{I}_{15}(x)$ are
hard to handle with this toolbox, we use the additional property that the
missing integrals satisfy a
hierarchically ordered coupled system of differential equations. This
particular property is
induced by the underlying sector
decomposition of the IBP method. More precisely, we are given
\begin{equation}\label{Equ:CoupledDESystem}
\begin{split}
D_x \hat{I}_{1}(x) =&-\tfrac{(-\ep+x-1)}{(x-1) x}\hat{I}_{1}(x)-\tfrac{2}{(x-1)
x}\hat{I}_{2}(x)+\tfrac{1}{(x-1) x}\hat{B}_{1}(x)+\dots\\
D_x \hat{I}_{2}(x)=&-\tfrac{\ep (3 \ep+2) (x-2)}{4 (x-1) x}\hat{I}_1(x)
+\tfrac{(-2+x+\ep (3 x-5))}{2 (x-1) x}\hat{I}_2(x)
-\tfrac{(2 \ep+x-\ep x)}{2 (x-1) x}\hat{I}_3(x)\\
&+\tfrac{\ep (50-14 x)+\ep^2 (25-6
x)-8(x-3)}{4(5 \ep+6) (x-1) x}\hat{B}_1(x)+\dots\\
D_x \hat{I}_{3}(x)=&\tfrac{\ep (3 \ep+2)}{4 (x-1)}\hat{I}_1(x)
+\tfrac{(2+\ep-3 x-3 \ep x)}{2 (x-1) x}\hat{I}_2(x)
-\tfrac{(\ep+1)}{2 (x-1)}\hat{I}_3(x)+\tfrac{8 (x-3)+\ep^2 (6 x-25)+2 \ep
(7x-25)}{4(5 \ep+6) (x-1) x}\hat{B}_1(x)
\\
\end{split}
\end{equation}
in terms of $\hat{I}_1(x),\hat{I}_2(x),\hat{I}_3(x)$. Note that by the
internal structure the right hand sides
of~\eqref{Equ:CoupledDESystem} are free of
$\hat{I}_{4}(x),\dots,\hat{I}_{15}(x)$. In the following subsections we will
demonstrate how this system can be solved in
$\hat{I}_1(x),\hat{I}_2(x),\hat{I}_3(x)$ by using the explicitly given
expansions of the $\hat{B}_i$ that we calculated already as a preprocessing
step.

Given this result, we will then turn to the remaining
$\hat{I}_i(x)$ with $i>3$. Here the hierarchical nature proceeds.
Given the
$\ep$-expansion of the $\hat{B}_i(x)$ and
$\hat{I}_1(x),\hat{I}_2(x),\hat{I}_3(x)$ in terms of indefinite nested sums and
products, we obtain a coupled system in terms of the unknowns
$\{\hat{I}_4(x),\hat{I}_5(x)\}$ and we can solve them again in terms of
indefinite nested sums and products. Summarizing, we continue iteratively, and
obtain closed forms for the clustered integrals in the given hierarchically
structured order:
\begin{multline}\label{Equ:Cluster}
\{\hat{I}_1(x),\hat{I}_2(x),\hat{I}_3(x)\}\to\{\hat{I}_4(x),\hat{I}_5(x)\}
\to\{\hat{I}_6(x),\hat{I}_7(x),\hat{I} _8(x)\}\\
\to
\{\hat{I}_9(x),\hat{I}_{10}(x)\}\to
\{\hat{I}_{11}(x),\hat{I}_{12}(x),\hat{I}_{13}(x)\}\to\{\hat{I}_{14}(x)\}\to\{
\hat {I}_{15}(x)\}.
\end{multline}

\subsection{Step 1: Transformation to a coupled recurrence system}

In order to solve the coupled system~\eqref{Equ:CoupledDESystem}, we first
derive a coupled system of difference equations that determines the
coefficients 
$I_1(N),I_2(N),I_3(N)$ of the power series
$$\hat{I}_i(x)=\sum_{N=0}^{\infty}I_i(N).$$
Namely, plugging in these generating functions into the first equation
of~\eqref{Equ:CoupledDESystem} yields
\begin{equation*}
D_x \sum_{N=0}^{\infty}I_{1}(N)x^N =-\tfrac{(-\ep+x-1)}{(x-1)
x}\sum_{N=0}^{\infty}I_{1}(N)x^N-\tfrac{2}{(x-1)
x}\sum_{N=0}^{\infty}I_{2}(N)x^N +\tfrac{1}{(x-1)
x}\sum_{N=0}^{\infty}B_1(N)x^N+\dots
\end{equation*}
Then applying $D_x$ on the summands of the power series and doing coefficient
comparison on both sides leads to
$$ N I_{1}(N-1)-(\ep+N+1) I_{1}(N)+2 I_{2}(N)=B_{1}(N)+\dots,$$
where on the right hand sides only the master integrals $B_i(N)$ (but not
$I_i$) arise.
Similarly, we apply this transformation to the other equations
in~\eqref{Equ:CoupledDESystem} and inserting the already computed
$\ep$- expansions of $B_i(N)$ we obtain the coupled recurrence system
\small
\begin{equation}\label{Equ:CoupledRESystem}
\begin{split}
 N I_{1}(N-1)&-(\ep+N+1) I_{1}(N)+2 I_{2}(N)\\ 
=&-\tfrac{4
(N+2)}{3(N+1)}\ep^{-3}+\Big(\tfrac{2 (2 N+1)}{3 (N+1)}S_1(N)-\tfrac{2 \big(6
N^2+13 N+8\big)}{3 (N+1)^2}\Big)\ep^{-2}+\dots \\
4(\ep-N) &I_{3}(N)-2 \ep (3 \ep+2) I_{1}(N)+\ep (3 \ep+2) I_{1}(N-1)\\
&-2 (3 \ep+1) I_{2}(N-1)+2 (5 \ep+2) I_{2}(N)-2 (\ep-2 N+1) I_{3}(N-1)\\
=& -\tfrac{8}{3}\ep^{-3}-\Big(\tfrac{8}{3}
S_1(N)-4\Big)\ep^{-2}-\Big(\tfrac{4}{3} S_1(N)^2-\tfrac{4
(N+1)}{N}S_1(N)+\tfrac{4}{3}S_2(N)+\zeta_2+6\Big)\ep^{-1}+\dots\\
2 (\ep+2 &N+2)I_{2}(N)-2 (3 \ep+2 N+1) I_{2}(N-1)+\ep (3 \ep+2) I_{1}(N-1)-2
(\ep+1) I_{3}(N-1)\\
=&\tfrac{8}{3}\ep^{-3}+\Big(\tfrac{8}{3}
S_1(N)-4\Big)\ep^{-2}+\Big(\tfrac{4}{3}
S_1(N)^2-\tfrac{4
(N+1)}{N}S_1(N)+\tfrac{4}{3}S_2(N)+\zeta_2+6\Big)\ep^{-1}+\dots, \\
\end{split}
\end{equation}
\normalsize
where the left hand sides contain the unknowns $I_1(N),I_2(N),I_3(N)$ and the
right hand sides consist of $\ep$-expansions whose coefficients are given in
terms of indefinite nested sums and products. More precisely, in our
concrete example, only harmonic sums occur.

\subsection{Step 2: Uncouple the recurrence system}

In the next step we uncouple the system~\eqref{Equ:CoupledRESystem} in the
following sense: we search for one scalar linear recurrence in one of the
functions, say $I_1(N)$, and express the remaining functions $I_2(N)$ and
$I_3(N)$ in terms of $I_1(N)$. To accomplish this task, various algorithms
are available within the Mathematica package~\texttt{OreSys}~\cite{Gerhold:02};
for our concrete problem we
took Z\"urcher's uncoupling algorithm~\cite{Zuercher:94}.

More precisely, we get the scalar difference equation
\begin{multline}\label{Equ:ScalarEq}
-2 (N+1) (N+2) (\ep+N+2) I_{1}(N)-(N+2) \big(2 \ep^2-5 \ep N-7 \ep-6 N^2-28
N-32\big) I_{1}(N+1)\\
+\big(\ep^3+4 \ep^2 N+14 \ep^2-4 \ep N^2-13 \ep N-3 \ep-6 N^3-50 N^2-136
N-120\big) I_{1}(N+2)\\
-(\ep-N-2) (\ep+N+4) (\ep+2 N+8) I_{1}(N+3)\\
=-\tfrac{4 (N+2)}{3 (N+3)}{\ep^{-3}}+\tfrac{2 \big(4 N^4+35 N^3+101 N^2+105
N+25\big)}{3 (N+1) (N+2) (N+3)^2}\ep^{-2}+\dots
\end{multline}
\normalsize
in the unknown function $I_1(N)$ and the two equations
\begin{equation}\label{Equ:DeterminRemainingIs}
\begin{split}
I_{2}(N)=&\square I_{1}(N)+\square I_{1}(N+1)+\square I_{1}(N+2)\\
&-\tfrac{2 (N+2)}{3 (N+1)}\ep^{-3}+\Big(\tfrac{6 N^3+25 N^2+33 N+15}{3 (N+1)^2
(N+2)}+\tfrac{(-2 N-1)}{3 (N+1)}S_1(N)\Big)\ep^{-2}+\dots\\[0.1cm]
I_{3}(N)=&\square I_{1}(N)+\square I_{1}(N+1)+\square I_{1}(N+2)\\
&+\tfrac{2 (N+2)}{3(N+1)}\ep^{-3}+\Big(\tfrac{-2 N^3-3 N^2+3 N+3}{3 (N+1)^2
(N+2)}+\tfrac{(2 N+1)}{3 (N+1)}S_1(N)\Big)\ep^{-2}+\dots
\end{split}
\end{equation}
that determine $I_2(N)$ and $I_3(N)$ if one knows the solution of $I_1(N)$.

\subsection{Step 3: Solve the uncoupled system of difference equations}

Now we are in the right position to activate Tactic 2 of our
symbolic summation toolbox. First we derive
the initial values
\begin{align*}
I_{1}(1)=&\!\!\tfrac{5}{\ep^3}\!-\!\tfrac{163}{12 \ep^2}+\big(\tfrac{15
\zeta_2}{8}\!+\!\tfrac{1223}{48}\big)\ep^{-1}+\dots,&
I_{1}(2)=&\!\!\tfrac{130}{27 \ep^3}\!-\!\tfrac{695}{54 \ep^2}\!+\!\big(\tfrac{65
\zeta_2}{36}+\tfrac{46379}{1944}\big)\ep^{-1}+\dots,\\
I_{1}(3)=&\!\!\frac{169}{36 \ep^3}\!-\!\frac{395}{32 \ep^2}\!+\!\big(\tfrac{169
\zeta_2}{96}\!+\!\tfrac{470071}{20736}\big)\ep^{-1}+\dots
\end{align*}
using, e.g., \texttt{MATAD}~\cite{Steinhauser:2000ry} or using further tools
as Mellin-Barnes integrals and other methods as
worked out
in~\cite{Ablinger:2014uka}.
Namely,
given~\eqref{Equ:ScalarEq} we activate \texttt{Sigma}'s recurrence solver and
obtain the $\ep$-expansion
\begin{align*}
I_{1}(N)&=\Big(\tfrac{4 \big(3 N^2+6 N+4\big)}{3 (N+1)^2}+\tfrac{4 S_1(N)}{3
(N+1)}\Big)\ep^{-3}\\
&+\Big(\tfrac{-2 \big(20 N^3+58 N^2+57 N+22\big)}{3
(N+1)^3}-\tfrac{S_1(N)^2}{N+1}+\tfrac{2 (N+2) (2 N-1) S_1(N)}{3
(N+1)^2}-\tfrac{S_2(N)}{N+1}\Big)\ep^{-2}+\dots
\end{align*}
where the coefficients are given in terms of harmonic sums. Finally, we
utilize~\eqref{Equ:DeterminRemainingIs} and get
\begin{align*}
I_{2}&(N)=\frac{4}{3}\ep^{-3}-\frac{2}{\ep^2}+\Big(-\tfrac{1}{3}
S_1(N)^2+\tfrac{2
}{3}S_1(N)-\tfrac{1}{3}S_2(N)+\tfrac{5 N+7}{3
(N+1)}+\tfrac{\zeta_2}{2}\Big)\ep^{-1}+\dots\\
I_{3}&(N)=-\tfrac{8}{3 \ep^3}+\Big(\tfrac{4 (N+2)}{3 (N+1)}S_1(N)-\tfrac{4
\big(4
N^2+7 N+2\big)}{3 (N+1)^2}\Big)\ep^{-2}\\
&+\Big(\tfrac{2 \big(12
N^3+32 N^2+25 N+2\big)}{3
(N+1)^3}-\tfrac{2 \big(4 N^2+11 N+10\big)}{3
(N+1)^2}S_1(N)+\tfrac{(N-2)}{3(N+1)}S_1(N)^2+\tfrac{(N-2)}{3
(N+1)}S_2(N)+\zeta_2\Big)\ep^{-1}+\dots
\end{align*}

\subsection{The general method and the physical result of $D_4(N)$}

Summarizing, we calculated the first coefficients of the $\ep$-expansions of
$\hat{I}_1(N), \hat{I}_2(N),\hat{I}_3(N)$ (resp.\
of $I_1(N),I_2(N),I_3(N)$) and treat also all other integrals
in~\eqref{Equ:Cluster} iteratively by the following method.

\medskip

\noindent\textit{Step 1:} Transform the coupled DE system to a coupled REC
System ($x\to N$).\\
\textit{Step 2:} Uncouple the REC system to a scalar recurrence for
one unknown integral, say $I_i(N)$.\\
\textit{Step 3:} Determine the coefficients of the $\ep$-expansion of $I_i(N)$
in terms of indefinite nested sums and products (see Tactic 2 in
Section~\ref{Sec:SymbolicSummation}) and derive the $\ep$-expansions of the
remaining integrals.\\ 
\textit{Step 4:} Translate back to the $x$-space
by $\hat{I}_i(x)=\sum_{N=0}^{\infty} I_i(N)x^N$.

\medskip 

To this end, we plug in all the computed expansions $\hat{B}_i(x)$ (by using
symbolic summation) and $\hat{I}_i(x)$ (by using our new solver for coupled
equations) into the expression~\eqref{Equ:D4XSpace}. This actually gives again
a gigantic expression in terms of generating functions where the coefficients
of the $\ep$-expansion are extremely large. In order to derive the $N$th
coefficient $D_4(N)$, we activate again our toolbox
mentioned in the introduction. We crunch the arising generating functions
with the package
\texttt{SumProduction} and compute the $N$th coefficient of the obtained
compact expression using the package \texttt{HarmonicSums}. Finally,
observe that this operation is based on Cauchy-product and we therefore obtain
again definite sums. Finally, we apply once more Tactic 1 of our symbolic
summation
toolbox, more precisely we use the package \texttt{EvaluateMultiSums} based on
\texttt{Sigma} to transform these sums to indefinite nested sums and products.

Summarizing, using all these packages,
we end up at the following result for~\eqref{Equ:Diagram}:
\begin{align*}
D_4(N)=&
\big(\frac{64 \big(N^2+N-1\big)}{3 (N+1) (N+2) (N+3) (N+4)}-\frac{64 S_1(N)}{3
(N+3) (N+4)}\Big)\ep^{-3}\\[0.1cm]
&+
\Big(
\frac{4 (N+1) (4 N+17) S_2(N)}{3
(N+2) (N+3)
(N+4)}-\frac{4 \big(3 N^5+68 N^4+379
N^3+648 N^2-98 N-696\big)}{3 (N+1) (N+2)^2 (N+3)^2 (N+4)^2}S_1(N)\\[-0.1cm]
&+\frac{4 \big(14 N^6+214 N^5+1179 N^4+3050 N^3+4097 N^2+3094 N+1200\big)}{3
(N+1)^2 (N+2)^2 (N+3)^2 (N+4)^2}\\
&+\frac{4 (5 N+27)}{3 (N+2) (N+3) (N+4)}S_1(N)^2
\Big)\ep^{-2}+\dots
\end{align*}
The single pole term and constant term are suppressed due to space
limitations. In total the following harmonic sums and generalized harmonic
sums
occur
\begin{align*}\quad\quad\quad&\zeta_2,\zeta_3,(-1)^N,2^N,S_{-3}(N),S_1(N),S_2(N)
,S_3(N),S_4(N),S_{-2,1}(N),S_{2,1}(N),S_{3,1}(N),\\
&S_1\big(\tfrac{1}{2},N\big),
S_1(2,N),
S_3\big(\tfrac{1}{2},N\big),S_{1,1}\big(1,\tfrac{1}{2},N\big),S_{1,1}\big(2,
\tfrac{1}{2},N\big),S_{2,1,1}(N),S_{2,1}\big(\tfrac{1}{2},1,
N\big),\\
&S_{2,1}\big(1,\tfrac{1}{2},N\big),S_{3,1}\big(\tfrac{1}{2},2,N\big),S_{1,1,1}
\big(1,1,\tfrac{1}{2},N\big),S_{2,1,1}\big(1,\tfrac{1}{2},2,N\big),S_{
1,1,1,1}\big(2,\tfrac{1}{2},1,1,N\big).
\end{align*}


\section{Conclusion}\label{Sec:Conclusion}

We presented a new method to solve coupled systems of differential
and difference equations which emerge in massive Feynman diagram
calculations. Here we rely on sophisticated summation tools based on difference
fields/rings and on uncoupling algorithms; for our concrete example we used
the
package
\texttt{OreSys}~\cite{Gerhold:02}. 

We obtained the $\epsilon$-expansions of rather complicated master
integrals. Using these expansions we calculated easily the most complicated
ladder graphs with 6 massive fermion lines using \texttt{Sigma}
\texttt{HarmonicSums}, \texttt{EvaluateMultiSums} and 
\texttt{SumProduction}. All ladder-topologies for 3-loop massive operator matrix
elements can be calculated in this way. The mass production is ready for graphs
depending on the same master integrals. We used this technology for a few
integrals emerging in Feynman integrals with two equal
masses~\cite{Ablinger:2014uka} and in the pure-singlet case \cite{Ablinger:14PS}. More
involved massive 3-loop topologies are currently investigated.


\begin{thebibliography}{99}
%
%
\bibitem{Ablinger:2010pb}
  J.~Ablinger, J.~Bl\"umlein, S.~Klein and C.~Schneider,
  {\it Modern Summation Methods and the Computation of 2- and 3-loop Feynman
  Diagrams},
  Nucl.\ Phys.\ Proc.\ Suppl.\  {\bf 205-206} (2010) 110--115
  [arXiv:1006.4797 [math-ph]].
%
\bibitem{Ablinger:2010ty}
  J.~Ablinger, J.~Bl\"umlein, S.~Klein, C.~Schneider and F.~Wi\ss{}brock,
  {\em The $O(\alpha_s^3)$ Massive Operator Matrix Elements of $O(N_f)$ for the
  Structure Function $F_2(x,Q^2)$ and 
  Transversity},
  Nucl.\ Phys.\ B {\bf 844} (2011) 26--54
  [arXiv:1008.3347 [hep-ph]].
%
\bibitem{Ablinger:2011te}
  J.~Ablinger, J.~Bl\"umlein and C.~Schneider,
  {\em Harmonic Sums and Polylogarithms Generated by Cyclotomic Polynomials},
  J.\ Math.\ Phys.\  {\bf 52} (2011) 102301
  [arXiv:1105.6063 [math-ph]].
%
\bibitem{Ablinger:2012qm}
  J.~Ablinger, J.~Bl\"umlein, A.~Hasselhuhn, S.~Klein, C.~Schneider and
  F.~Wi\ss{}brock,
  {\em Massive 3-loop Ladder Diagrams for Quarkonic Local Operator Matrix
  Elements},
  Nucl.\ Phys.\ B {\bf 864} (2012) 52--84
  [arXiv:1206.2252 [hep-ph]].
%
\bibitem{Ablinger:2012LL}
  J.~Ablinger, J.~Bl\"umlein, M.~Round and C.~Schneider,
  \textit{Advanced Computer Algebra Algorithms for the Expansion of Feynman
  Integrals},
  PoS LL {\bf 2012} (2012) 050
  [arXiv:1210.1685 [cs.SC]].
%
\bibitem{Ablinger:2013hcp}
  J.~Ablinger,
  \textit{Computer Algebra Algorithms for Special Functions in Particle
  Physics}, Ph.D. thesis, RISC, J. Kepler University Linz, 2013
  arXiv:1305.0687 [math-ph].
%
\bibitem{Ablinger:2013cf}
  J.~Ablinger, J.~Bl\"umlein and C.~Schneider,
  {\em Analytic and Algorithmic Aspects of Generalized Harmonic Sums and
  Polylogarithms},
  J.\ Math.\ Phys.\  {\bf 54} (2013) 082301
  [arXiv:1302.0378 [math-ph]].
%
\bibitem{Ablinger:2014vwa}
  J.~Ablinger, A.~Behring, J.~Bl\"umlein, A.~De Freitas, A.~Hasselhuhn, A.~von Manteuffel, M.~Round and 
  C.~Schneider and F.~Wi\ss{}brock,
  {\it The 3-Loop Non-Singlet Heavy Flavor Contributions and Anomalous Dimensions for the Structure Function 
  $F_2(x,Q^2)$ and Transversity},
  arXiv:1406.4654 [hep-ph].
%
\bibitem{Ablinger:2014lka}
  J.~Ablinger, J.~Bl\"umlein, A.~De Freitas, A.~Hasselhuhn, A.~von Manteuffel,
  M.~Round, C.~Schneider and F.~Wi\ss{}brock,
  {\em The Transition Matrix Element $A_{gq}(N)$ of the Variable Flavor Number
  Scheme at $O(\alpha_s^3)$},
  Nucl.\ Phys.\ B {\bf 882} (2014) 263--288
  [arXiv:1402.0359 [hep-ph]].
%
\bibitem{Ablinger:2014uka}
  J.~Ablinger, J.~Bl\"umlein, A.~De Freitas, A.~Hasselhuhn, A.~von Manteuffel,
  M.~Round and C.~Schneider,
  {\em The $O(\alpha_s^3 T_F^2)$ Contributions to the Gluonic Operator Matrix
  Element}
  Nucl. Phys. B {\bf 885} (2014) 280--317 [arXiv:1405.4259 [hep-ph]].
%
\bibitem{Ablinger:2014bra}
  J.~Ablinger, J.~Bl\"umlein, C.~G.~Raab and C.~Schneider,
  {\it Iterated Binomial Sums and their Associated Iterated Integrals},
  arXiv:1407.1822 [hep-th].
%
\bibitem{Ablinger:2014yaa}
  J.~Ablinger, J.~Bl\"umlein, C.~Raab, C.~Schneider and F.~Wi\ss{}brock,
  {\em Calculating Massive 3-loop Graphs for Operator Matrix Elements by the
  Method of Hyperlogarithms},
  Nucl. Phys.  B {\bf 885} (2014) 409--447 [arXiv:1403.1137 [hep-ph]].
%
\bibitem{Ablinger:14PS}
J.~Ablinger et al., DESY 13--232.
%
\bibitem{Abramov:94}
S.A.~Abramov and M.~Petkov{\v s}ek, \textit{D'{A}lembertian solutions of linear
  differential and difference equations},
In: J.~von~zur Gathen (ed.) Proc. ISSAC'94, pp. 169--174 ACM Press (1994).
%
\bibitem{Bierenbaum:2008yu}
  I.~Bierenbaum, J.~Bl\"umlein, S.~Klein and C.~Schneider,
  {\em Two-Loop Massive Operator Matrix Elements for Unpolarized Heavy Flavor
  Production to $O(\varepsilon)$},
  Nucl.\ Phys.\ B {\bf 803} (2008) 1--41
  [arXiv:0803.0273 [hep-ph]].
%
\bibitem{Blumlein:1998if}
  J.~Bl\"umlein and S.~Kurth,
  {\em Harmonic sums and Mellin transforms up to two loop order},
  {Phys.\ Rev.}\ D {\bf 60} (1999) 014018
  [arXiv:9810241 [hep-ph]].
%
\bibitem{Blumlein:2010zv}
  J.~Bl\"umlein, S.~Klein, C.~Schneider and F.~Stan,
  {\it A Symbolic Summation Approach to Feynman Integral Calculus},
  J. Symbolic Comput. 47 (2012) 1267--1289
  [arXiv:1011.2656 [cs.SC]].
%
\bibitem{Blumlein:2012vq}
  J.~Bl\"umlein, A.~Hasselhuhn, S.~Klein and C.~Schneider,
  {\em The $O(\alpha_s^3 N_f T_F^2 C_{A,F})$ Contributions to the Gluonic
  Massive Operator Matrix Elements},
  Nucl.\ Phys.\ B {\bf 866} (2013) 196--211
  [arXiv:1205.4184 [hep-ph]].
%
\bibitem{Blumlein:2012hg}
  J.~Bl\"umlein, A.~Hasselhuhn and C.~Schneider,
  {\it Evaluation of Multi-Sums for Large Scale Problems},
  PoS RADCOR {\bf 2011} (2011) 032
  [arXiv:1202.4303 [math-ph]].
%
\bibitem{Bron:00}
M. Bronstein, \textit{On solutions of linear ordinary difference equations in
their coefficient field}, J.~Symbolic Comput. \textbf{29} (2000) 841--877.
%
\bibitem{Brown:2008um}
  F.C.S.~Brown,
  \textit{The Massless higher-loop two-point function},
  Commun.\ Math.\ Phys.\  {\bf 287} (2009) 925--958
  [arXiv:0804.1660 [math.AG]].
%
\bibitem{Chetyrkin:1981qh}
  K.~G.~Chetyrkin and F.~V.~Tkachov,
  {\it Integration by Parts: The Algorithm to Calculate beta Functions in 4
  Loops},
  Nucl.\ Phys.\ B {\bf 192} (1981) 159--204
%
\bibitem{Davydychev:2003mv}
  A.~I.~Davydychev and M.~Y.~Kalmykov,
  {\it Massive Feynman diagrams and inverse binomial sums},
  Nucl.\ Phys.\ B {\bf 699} (2004) 3--64 [arXiv:0303162 [hep-th]].
%
\bibitem{Fleischer:1998nb}
  J.~Fleischer, A.~V.~Kotikov and O.~L.~Veretin,
  {\it Analytic two loop results for selfenergy type and vertex type diagrams with one nonzero mass},
  Nucl.\ Phys.\ B {\bf 547} (1999) 343--374
  [hep-ph/9808242].
%
\bibitem{Gerhold:02}
S.~Gerhold, {\it Uncoupling systems of linear {O}re operator equations},
Master's thesis, RISC, J.~Kepler University, Linz, 2002.
%
\bibitem{Karr:81}
 M.~Karr,
 \newblock {\em Summation in finite terms},
 \newblock {J. ACM}, {\bf 28} (1981) 305--350.
%
\bibitem{Laporta:1996mq}
  S.~Laporta and E.~Remiddi,
  {\it The Analytical value of the electron $(g-2)$ at order $\alpha^3$ in QED},
  Phys.\ Lett.\ B {\bf 379} (1996) 283--291
  [arXiv:9602417 [hep-ph]].
%
\bibitem{Laporta:2001dd}
  S.~Laporta,
  {\it High precision calculation of multiloop Feynman integrals by difference equations},
  Int.\ J.\ Mod.\ Phys.\ A {\bf 15} (2000) 5087--5159
  [arXiv:0102033 [hep-ph]].
%
\bibitem{vonManteuffel:2012np}
  A.~von Manteuffel and C.~Studerus,
  {\it Reduze\!~\!2 - Distributed Feynman Integral Reduction}, 2012
  [arXiv:1201.4330 [hep-ph]].
%
\bibitem{Moch:2001zr}
  S.~Moch, P.~Uwer and S.~Weinzierl,
  {\em Nested sums, expansion of transcendental functions and multiscale multiloop integrals},
  J.\ Math.\ Phys.\  {\bf 43} (2002) 3363--3386
  [arXiv:0110083 [hep-ph]].
%
\bibitem{Petkov:92}
M.~Petkov{\v s}ek, \textit{Hypergeometric solutions of linear recurrences with polynomial coefficients},
J.~Symbolic Comput. \textbf{14} (1992) 243--264.
%
\bibitem{Schneider:01}
C.~Schneider,  
{\it Symbolic Summation in Difference Fields\/}, Ph.D. Thesis
RISC, Johannes Kepler University, Linz technical report 01-17 (2001).
%
\bibitem{Schneider:05a}
C.~Schneider, \textit{Solving parameterized linear difference equations in terms
of indefinite nested sums and products}, J. Differ. Equations Appl. {\bf 11}
(2005) 799--821.
%
\bibitem{SIG1}
C.~Schneider, 
{\em Symbolic summation assists combinatorics},
{S\'em.~Lothar. Combin.\/} {\bf 56} (2007) 1--36
 article B56b.
%
\bibitem{Schneider:08c}
C.~Schneider, \textit{ A refined difference field theory for symbolic
summation}, {J. Symbolic Comput.\/} {\bf 43} (2008) 611--644
  [arXiv:0808.2543 [cs.SC]].
%
\bibitem{Schneider:10a}
C.~Schneider, \textit{Structural Theorems for Symbolic Summation}, {Appl.
Algebra Engrg. Comm. Comput.} {\bf 21} (2010) 1--32.
%
\bibitem{Schneider:10b}
C.~Schneider, \textit{A Symbolic Summation Approach to Find Optimal Nested Sum
  Representations}, 
\newblock In A.~Carey, D.~Ellwood, S.~Paycha, and S.~Rosenberg, editors, {\sf
  {Motives, Quantum Field Theory, and Pseudodifferential Operators}}, Vol.~{\bf
12}
  {Clay Mathematics Proceedings},
  Amer. Math. Soc., pp.~285--308, (2010)
  [arXiv:0808.2543 [cs.SC]].
%
\bibitem{Schneider:10c}
C.~Schneider, \textit{Parameterized Telescoping Proves Algebraic Independence of
  Sums}, Ann. Comb. {\bf 14} (2010)  533--552 
[arXiv:0808.2596 [cs.SC]].
%
\bibitem{SIG2}
C.~Schneider, {\it Simplifying Multiple Sums in Difference Fields}, in:~{ 
Computer Algebra in Quantum Field Theory: Integration,
  Summation and Special Functions}, Texts \&  Monographs in Symbolic
  Computation eds. C.~Schneider and J.~Bl\"umlein  (Springer, Wien, 2013) 325--360
  [arXiv:1304.4134 [cs.SC]].
%
\bibitem{Schneider:13b}
C.~Schneider, \textit{Fast Algorithms for Refined Parameterized Telescoping in
Difference Fields},
in~: Lecture Notes in Computer Science (LNCS)
eds. J. Guitierrez, J. Schicho, M. Weimann, in press (2014) [arXiv:13077887 [cs.SC]].
%
\bibitem{Schneider:2013zna}
  C.~Schneider,
  {\it Modern Summation Methods for Loop Integrals in Quantum Field Theory: The
Packages Sigma, EvaluateMultiSums and 
  SumProduction},
  J.\ Phys.\ Conf.\ Ser.\  {\bf 523} (2014) 012037
  [arXiv:1310.0160 [cs.SC]].
%
\bibitem{Steinhauser:2000ry}
  M.~Steinhauser,
  {{\tt MATAD:} A Program package for the computation of MAssive TADpoles},
  Comput.\ Phys.\ Commun.\  {\bf 134} (2001) 335--364
  [hep-ph/0009029].
%
\bibitem{Studerus:2009ye}
  C.~Studerus,
  {\it Reduze-Feynman Integral Reduction in {\tt C++}},
  Comput.\ Phys.\ Commun.\  {\bf 181} (2010) 1293
  [arXiv:0912.2546 [physics.comp-ph]].
%
\bibitem{Tkachov:1981wb}
  F.~V.~Tkachov,
  {\it A Theorem on Analytical Calculability of Four Loop Renormalization Group
Functions},
  Phys.\ Lett.\ B {\bf 100} (1981) 65--68.
%
\bibitem{Vermaseren:1994je}
  J.A.M.~Vermaseren,
  {\it Axodraw}, 
  Comput.\ Phys.\ Commun.\  {\bf 83} (1994) 45--58.
%
\bibitem{Vermaseren:1998uu}
  J.A.M.~Vermaseren,
  {\em Harmonic sums, Mellin transforms and integrals},
  Int.\ J.\ Mod.\ Phys.\ A {\bf 14} (1999) 2037--2076
  [arXiv: 9806280 [hep-ph]].
%
\bibitem{Weinzierl:2004bn}
  S.~Weinzierl,
  {\it Expansion around half integer values, binomial sums and inverse binomial
sums},
  J.\ Math.\ Phys.\  {\bf 45} (2004) 2656--2673
  [arXiv:0402131 [hep-ph]].
%
\bibitem{Weinzierl:13}
S.~Weinzierl, {\it Feynman graphs}, 
in:~{Computer Algebra in Quantum Field Theory: Integration,
  Summation and Special Functions}, Texts \&  Monographs in Symbolic
  Computation eds. C.~Schneider and J.~Bl\"umlein  (Springer, Wien, 2013)
  381--406 (2013) [arXiv:13016918 [hep-ph]].
%
\bibitem{Zeilberger:91}
D.~Zeilberger, {\it The method of creative telescoping},
J.~Symbolic Comput. \textbf{11}, 195--204 (1991)
%
\bibitem{Zuercher:94}
B.~Z\"urcher, {\it Rationale Normalformen von pseudo-linearen Abbildungen},
Master's thesis, Mathematik, ETH Z\"urich, 1994.
\end{thebibliography}
\end{document}